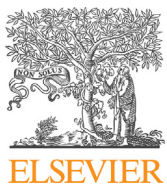
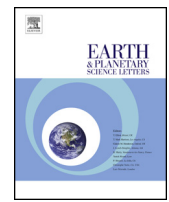

Contents lists available at ScienceDirect

# Earth and Planetary Science Letters

www.elsevier.com/locate/epsl

# Late metal–silicate separation on the IAB parent asteroid: Constraints from combined W and Pt isotopes and thermal modelling

Alison C. Hunt [a,∗], David L. Cook [a], Tim Lichtenberg [b,c], Philip M. Reger [a], Mattias Ek [a], Gregor J. Golabek [d], Maria Schönbächler [a]

[a] *Institute of Geochemistry and Petrology, ETH Zürich, Clausiusstrasse 25, 8092 Zürich, Switzerland*
[b] *Institute of Geophysics, ETH Zürich, Sonneggstrasse 5, 8092 Zürich, Switzerland*
[c] *Institute for Astronomy, ETH Zürich, Wolfgang-Pauli-Strasse 27, 8093 Zürich, Switzerland*
[d] *Bayerisches Geoinstitut, University of Bayreuth, Universitätsstrasse 30, 95440 Bayreuth, Germany*

### a b s t r a c t

The short-lived $^{182}$Hf–$^{182}$W decay system is a powerful chronometer for constraining the timing of metal–silicate separation and core formation in planetesimals and planets. Neutron capture effects on W isotopes, however, significantly hamper the application of this tool. In order to correct for neutron capture effects, Pt isotopes have emerged as a reliable in-situ neutron dosimeter. This study applies this method to IAB iron meteorites, in order to constrain the timing of metal segregation on the IAB parent body.

The $\varepsilon^{182}$W values obtained for the IAB iron meteorites range from $-3.61 \pm 0.10$ to $-2.73 \pm 0.09$. Correlating $\varepsilon^{i}$Pt with $\varepsilon^{182}$W data yields a pre-neutron capture $\varepsilon^{182}$W of $-2.90 \pm 0.06$. This corresponds to a metal–silicate separation age of $6.0 \pm 0.8$ Ma after CAI for the IAB parent body, and is interpreted to represent a body-wide melting event. Later, between 10 and 14 Ma after CAI, an impact led to a catastrophic break-up and subsequent reassembly of the parent body. Thermal models of the interior evolution that are consistent with these estimates suggest that the IAB parent body underwent metal–silicate separation as a result of internal heating by short-lived radionuclides and accreted at around $1.4 \pm 0.1$ Ma after CAIs with a radius of greater than 60 km.

## 1. Introduction

The short-lived $^{182}$Hf–$^{182}$W chronometer ($t_{1/2} = 8.9$ Ma) is a valuable tool that can effectively constrain the timing of metal–silicate separation and core formation in planetary bodies. Hafnium and W are refractory elements that are strongly fractionated from each other during this process. Hafnium is lithophile and partitions into the silicate phase, whereas W is siderophile and partitions into metallic phases. The application of the $^{182}$Hf–$^{182}$W chronometer to solar system materials, however, is hindered by the effects of neutron capture on W isotopes during exposure to galactic cosmic rays (GCR) (e.g., Leya and Masarik, 2013). Neutron capture reactions cause burnout of $^{182}$W, leading to incorrect $^{182}$Hf–$^{182}$W ages. Platinum is an ideal in-situ neutron dose proxy due to the similar neutron capture cross-sections of Pt and W (Leya and Masarik, 2013). Correlating the effects in Pt and W isotopic compositions allows for an estimation of the pre-GCR exposure $\varepsilon^{182}$W value, and consequently, the timing of metal–silicate separation. This method was successfully applied to magmatic iron meteorites by several studies (Kruijer et al., 2014a, 2013a; Wittig et al., 2013).

The IAB meteorites are non-magmatic iron meteorites that have been classified into numerous sub-groups based on their trace-element geochemistry. These include the Main Group (MG), multiple sub-groups including 'Low-Au Low-Ni' (sLL) and 'Low-Au Medium-Ni' (sLM), and several duos or ungrouped samples (Wasson and Kallemeyn, 2002). Recently, nucleosynthetic Mo isotope variations have been used to identify genetic links between these sub-groups (Worsham et al., 2017). That study suggests that while the MG, sLL, sLH and sLM are genetically related, two high-Au sub-groups (sHL and sHH), as well as some ungrouped samples, may derive from a distinct parent body or parent body family.

Trace element data, including highly variable concentrations of siderophile elements, suggest the IABs did not form by simple fractional crystallization in a planetary core, in contrast to magmatic irons (e.g., Wasson and Kallemeyn, 2002; Worsham et al., 2016). Additionally, the IAB irons contain inclusions of chondritic and non-chondritic silicates, sulfides, graphite, and phosphate-bearing inclusions (e.g., Benedix et al., 2000; Buchwald, 1975). The angular texture and the olivine- and pyroxene-rich mineralogy of the chondritic silicate inclusions are similar to the winonaite primitive achondrites (Benedix et al., 1998; Bild, 1977). A genetic relationship between the winonaites and MG-IAB iron meteorites is also implied by the identical Mo and O isotope compositions of the two groups, suggesting that they originated from a common parent body (Clayton and Mayeda, 1996; Greenwood et al., 2012; Worsham et al., 2017).

\* Corresponding author.
*E-mail address:* alison.hunt@erdw.ethz.ch (A.C. Hunt).

**Table 1**
Sample information and Ir/Pt ratios for analysed IAB meteorites.

| Samples | Source | IAB group/subgroup[a] | Description of inclusions[b] | Ir/Pt[a] | CRE ages (Ma) |
|---|---|---|---|---|---|
| Caddo County | ETH | un | ac, nc | 0.40 | 5 ± 1[c] |
| Canyon Diablo | ETH | MG | gr, ac | 0.40 | 545 ± 40[d] |
| Cranbourne | ETH | MG | gr | 0.34 | 45 ± 5[e] |
| Livingstone (TN) | Smithsonian (USNM 1420) | Algarrabo duo | gr | 0.17 | |
| Magura | ETH | MG | gr, s | 0.37 | ∼250[f] |
| Odessa | ETH | MG | gr | 0.41 | 875 ± 70[g] |
| Toluca | ETH | sLL | gr, s | 0.43 | 600 ± 150[e] |

[a] IAB classifications and Ir/Pt ratios from Wasson and Kallemeyn (2002). IAB group/sub-groups: MG, main group; sLL, subgroup low-Au low-Ni; un, ungrouped.
[b] Petrographic descriptions of inclusions: ac, angular chondritic; nc, non-chondritic; gr, graphite-rich; s, silicate (Benedix et al., 2000; Buchwald, 1975).
[c] CRE ages from: Takeda et al. (2000).
[d] CRE ages from: Schnabel et al. (2001).
[e] CRE ages from: Chang and Wänke (1969).
[f] CRE ages from: Schulz et al. (2012).
[g] CRE ages from: Voshage and Feldmann (1979).

The thermal history and evolution of the IAB parent asteroid, or asteroid family, remains unclear. One theory suggests that the IAB irons formed in melt pools created by impacts into a chondritic parent body or bodies (Choi et al., 1995; Wasson and Kallemeyn, 2002; Worsham et al., 2017, 2016). A competing theory argues that the decay of short-lived radionuclides such as $^{26}$Al and $^{60}$Fe would produce enough heat for incipient partial melting, which was followed by a catastrophic impact. Subsequent reassembly due to gravity led to the extensive mixing of silicates, sulfides and metals (Benedix et al., 2000; Hunt et al., 2017a; Schulz et al., 2012; Theis et al., 2013). This model can apply to groups with the same genetic affinities (e.g., the MG, sLL, sLH and sLM) as defined by Mo isotopes (Worsham et al., 2017).

Several studies attempted to constrain the history of the IAB parent body using the $^{182}$Hf–$^{182}$W chronometer (e.g., Markowski et al., 2006; Qin et al., 2008; Schulz et al., 2009). The metal–silicate separation ages based on these data and recalculated relative to the most recent value for calcium aluminum rich inclusions (CAI) ($\varepsilon^{182}$W: −3.49 ± 0.07, Burkhardt and Schönbächler, 2015; Kruijer et al., 2014b) range from 3.6 ± 2.1 Ma (Schulz et al., 2009) to 11.5 ± 6.5 Ma after CAI (Markowski et al., 2006). This broad range is partly related to difficulties to correct for neutron capture effects, reasserting the need for a reliable neutron-dose proxy. Recently, attempts to correct for the effects of GCR have yielded $^{182}$Hf–$^{182}$W ages for the IAB irons of between 3.4 ± 0.7 and 6.9 ± 0.4 Ma (Schulz et al., 2012; Worsham et al., 2017). This study also corrects for the effect of GCR on W isotopes in the IABs, using the well-established Pt isotope neutron-dose proxy. We report new W and Pt isotope data obtained on the same sample aliquot for seven samples. The data provide new insights into the thermal evolution of the IAB parent body by constraining the timing of metal–silicate separation. Numerical models of parent body evolution supplement these results. Additionally, the potential of Pt isotopes as an in-situ neutron-dose proxy in iron meteorites is further evaluated, specifically for non-magmatic irons.

## 2. Samples

Seven IAB meteorites originating from four different IAB groups or sub-groups, as defined by Wasson and Kallemeyn (2002), were chosen for this study. Canyon Diablo, Cranbourne, Magura and Odessa are all assigned to the MG, whereas Toluca is a member of the sLL sub-group, Livingstone (TN) is part of the Algarrabo duo, and Caddo County is ungrouped, but a member of the Udei Station grouplet. These meteorites contain a variety of inclusion types, and cover a range of cosmic ray exposure (CRE) ages (Table 1). All samples are taken from the ETH collection, except Livingstone (TN) (USNM 1420, Smithsonian Institute).

## 3. Methods

### 3.1. High-precision Pt and W isotope measurements

Between 1.5 and 3.0 g of each IAB meteorite was prepared for use in this study, and both W and Pt isotope aliquots were taken from the same sample digestion. The sample preparation procedures and Pt and W isotope chemical separations employed in this study are described in the Supplementary Materials. Platinum isotope data for Toluca are taken from Hunt et al. (2017b; Toluca-a).

All Pt and W isotope analyses were performed at ETH Zürich using a Thermo Scientific Neptune *Plus* fitted with a Cetac Aridus II desolvating system and standard H cones. Platinum isotope analyses followed the procedure of Hunt et al. (2017c). Analyses were corrected for instrumental mass bias using the exponential law, and were internally normalized to $^{198}$Pt/$^{195}$Pt ('8/5') = 0.2145 (Kruijer et al., 2013a). Samples were analysed with a $^{194}$Pt signal of ∼2 × 10$^{−10}$ A, equivalent to ∼200 ppb Pt, and utilizing ∼200 ng Pt per measurement. Each sample was measured relative to the NIST SRM 3140 Pt standard solution, and data are presented in the epsilon notation (i.e., $\varepsilon^{19i}$Pt/$^{195}$Pt = deviation in parts per 10,000 from the average of the bracketing standards). Repeat analysis of four aliquots of our in-house reference material, the North Chile iron meteorite (IIAB), passed through column chemistry independently give a 2 standard deviation external reproducibility (2 S.D.) of 0.73 for $\varepsilon^{192}$Pt, 0.15 for $\varepsilon^{194}$Pt, and 0.09 for $\varepsilon^{196}$Pt ($n = 19$; Hunt et al., 2017c). Our 2 S.D. external reproducibility for repeat analyses of North Chile is used as the Pt isotope uncertainty throughout this study.

All five W isotopes, along with the interference monitors $^{178}$Hf, $^{181}$Ta, and $^{188}$Os, were measured simultaneously in static mode (Cook and Schönbächler, 2016). Single measurements of each sample were bracketed by measurements of the NIST SRM 3163 W

**Table 2**
Platinum and W isotope compositions of the IAB irons.

| | IAB group[a] | n (Pt) | $\varepsilon^{192}$Pt | ±2 S.D. | $\varepsilon^{194}$Pt | ±2 S.D. | $\varepsilon^{196}$Pt | ±2 S.D. | n (W) | $\varepsilon^{182}$W (6/4) | ±2 S.D. | $\varepsilon^{182}$W (6/3)[b] | ±2 S.D. | $\varepsilon^{184}$W (6/3)[b] | ±2 S.D. |
|---|---|---|---|---|---|---|---|---|---|---|---|---|---|---|---|
| Caddo County | un | 1 | 0.09 | 0.73 | 0.06 | 0.15 | −0.01 | 0.09 | 1 | −2.92 | 0.10 | −2.89 | 0.10 | 0.04 | 0.06 |
| Canyon Diablo | MG | 2 | −0.48 | 0.73 | 0.04 | 0.15 | 0.00 | 0.09 | 1 | −2.87 | 0.10 | −2.87 | 0.10 | 0.00 | 0.06 |
| Cranbourne | MG | 1 | 2.13 | 0.73 | 0.34 | 0.15 | 0.13 | 0.09 | 1 | −3.16 | 0.10 | −3.16 | 0.10 | 0.06 | 0.06 |
| Magura | MG | 2 | 2.08 | 0.73 | 0.21 | 0.15 | 0.11 | 0.09 | 1 | −3.11 | 0.10 | −3.11 | 0.10 | 0.01 | 0.06 |
| Odessa | MG | 3 | 12.99 | 0.73 | 0.74 | 0.15 | 0.56 | 0.09 | 1 | −3.61 | 0.10 | −3.61 | 0.10 | 0.02 | 0.06 |
| Livingston (TN) | Algarrabo duo | 2 | 0.03 | 0.73 | 0.24 | 0.15 | 0.08 | 0.09 | 1 | −3.12 | 0.09 | −3.17 | 0.07* | 0.04 | 0.07 |
| Toluca[c] | sLL | 1 | −0.17 | 0.73 | 0.09 | 0.15 | 0.00 | 0.09 | 1 | −2.73 | 0.09 | −2.73 | 0.07* | 0.00 | 0.07 |

All Pt isotope data are normalised to $^{198}$Pt/$^{195}$Pt = 0.2145 (Kruijer et al., 2013a). Platinum isotope uncertainties are based on repeat analysis of the in-house standard (North Chile IIAB iron meteorite; see section 3.1). Tungsten isotope compositions are normalized relative to $^{186}$W/$^{183}$W = 1.98594 (6/3) or $^{186}$W/$^{184}$W = 0.927672 (6/4). Uncertainties are based on the W bracketing standard (NIST SRM 3163) or the long-term reproducibility defined by repeated measurements of SRM 129c (denoted by an asterisk; Cook and Schönbächler, 2016), whichever is larger.
[a] See Table 1 for definition of groups.
[b] Tungsten isotope ratios corrected for nuclear field shift effects, as described in Cook and Schönbächler (2016). Uncorrected data are given in Table S1 of the Supplementary Appendix.
[c] Pt isotope data previously published as Toluca-a (Hunt et al., 2017b).

solution standard. Concentrations were adjusted to achieve a signal of $\sim 4.5 \times 10^{-10}$ A on $^{184}$W. Instrumental mass bias was corrected using the exponential law with either $^{186}$W/$^{183}$W = 1.98594 ('6/3') or $^{186}$W/$^{184}$W = 0.927672 ('6/4') (Völkening et al., 1991) and data are presented in the epsilon notation. Additionally, data are corrected for nuclear field shift (NFS) effects using the method outlined in Cook and Schönbächler (2016). Uncertainties on W isotope ratios are mainly based on the deviation (2 S.D.) of the bracketing standard analyses during each analytical session (Table 2). Five replicates of the NIST Fe–Ni steel, SRM 129c, were also analyzed in each session; this material was used as an external standard to define the long-term reproducibility (2 S.D.) of the isotopic measurements and to validate their accuracy during the analytical campaign. These data are published in Cook and Schönbächler (2016). Epsilon values based on SRM 129c have the following external precisions: ± 0.05 for $\varepsilon^{184}$W (6/3), ± 0.07 for $\varepsilon^{182}$W (6/3), and ± 0.08 for $\varepsilon^{182}$W (6/4) (Cook and Schönbächler, 2016). The external precision of SRM 129c is similar to that derived using the bracketing standard NIST SRM 3163. However, for each W isotope ratio presented, the larger of the two values is taken as the uncertainty (Table 2).

### 3.2. Numerical modelling

The thermal evolution of the IAB parent body was modelled using both 2D and 3D fluid dynamics simulations employing the I2/I3ELVIS code family (Gerya and Yuen, 2007, 2003). The code uses a finite-differences fully-staggered grid and solves for the conservation equations for mass, momentum and energy including material self-gravity. The numerical model accounts for heating by radiogenic, shear and latent heat production source terms. Each model was assigned an initial $^{26}$Al/$^{27}$Al ratio and planetesimal radius, which together determined the internal evolution. The $^{26}$Al/$^{27}$Al varied from $2.5 \times 10^{-5}$ to $0.7 \times 10^{-5}$ (corresponding to formation times $t_{form}$ = 0.75 to 2.0 Ma after CAI, using the assumption of an initially homogeneous $^{26}$Al distribution in the protoplanetary disk with the canonical initial $^{26}$Al/$^{27}$Al of $5.25 \times 10^{-5}$; Kita et al., 2013) and planetesimal radii $R_P$ from 20 to 200 km. The silicate melt fraction was linearly parameterized taking into account both consumption and release of latent heat. For models reaching high melt fractions above 0.4, the convective heat flux was approximated following soft turbulence scaling relations (Kraichnan, 1962; Siggia, 1994). The model incorporates an initial macroporosity of 0.3, with sintering and compaction effects parameterized according to laboratory experiments (Henke et al., 2012; Gail et al., 2015). An ambient and starting temperature of 290 K was employed. Cooling at the surface was handled via a low-density and low-viscosity layer of 'sticky air' surrounding the sil-icate body (Schmeling et al., 2008; Crameri et al., 2012). For an in-depth discussion of the simulation method see Golabek et al. (2014). The thermomechanical evolution regimes of early solar system planetesimals are described in Lichtenberg et al. (2016).

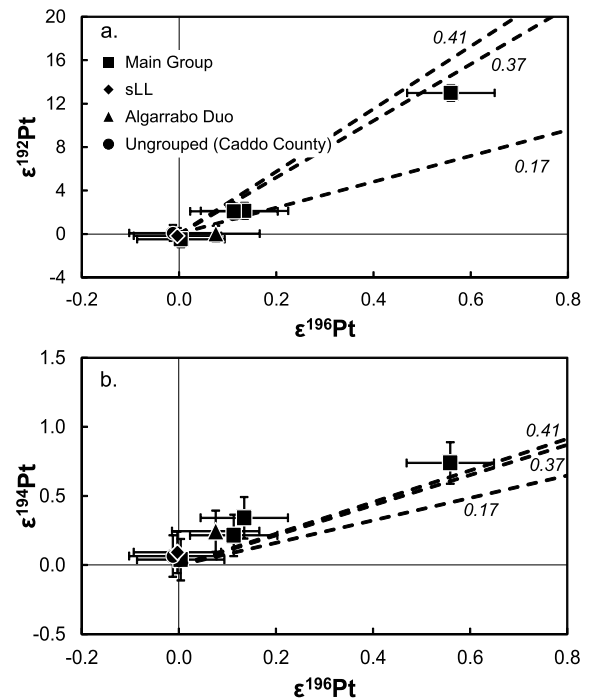

**Fig. 1.** Co-variation of $\varepsilon^{196}$Pt vs. (a) $\varepsilon^{192}$Pt and (b) $\varepsilon^{194}$Pt for the IAB samples. Also shown are the model calculations of Leya and Masarik (2013) for Ir/Pt ratios of 0.17 to 0.41, an exposure time of 1000 Ma and all shielding depths across pre-atmospheric radii of 5 to 120 cm (dashed lines). The model lines for the different Ir/Pt ratios illustrate the influence of varying Ir/Pt on $\varepsilon^{192}$Pt; see Table 1 for Ir/Pt ratios of the studied IABs. Uncertainties (2 S.D.) are based on the North Chile in-house standard.

### 4. Results

#### 4.1. Platinum isotope data

The three samples Caddo County, Canyon Diablo and Toluca display Pt isotope compositions that overlap within uncertainty with the terrestrial value (Table 2; Fig. 1). In contrast, Cranbourne and Magura both exhibit small but resolvable positive shifts ($\varepsilon^{192}$Pt = 2.13 and 2.08 ± 0.73; $\varepsilon^{196}$Pt = 0.13 and 0.11 ± 0.09, respectively), whereas Odessa shows large excesses (i.e., $\varepsilon^{192}$Pt = 12.99 ± 0.73;

$\varepsilon^{196}$Pt $= 0.56 \pm 0.09$) relative to the terrestrial standard. Livingstone falls within uncertainty of the terrestrial value in the $\varepsilon^{192}$Pt vs. $\varepsilon^{196}$Pt diagram, but not when considering $\varepsilon^{194}$Pt vs. $\varepsilon^{196}$Pt space.

### 4.2. Tungsten isotope data

The IAB irons show variable W isotope compositions (Table 2). Toluca displays the least negative $\varepsilon^{182}$W (6/4) value among the IABs ($-2.73 \pm 0.09$). The $\varepsilon^{182}$W (6/4) values of Caddo County and Canyon Diablo are $-2.92$ and $-2.87 \pm 0.10$, respectively, whereas Cranbourne, Livingstone and Magura yield identical values within uncertainty (Table 2). Odessa displays the most negative $\varepsilon^{182}$W ($-3.61 \pm 0.10$). Tungsten isotope ratios may be affected by NFS effects that arise during chemical separation of W, particularly ratios that include and/or are normalized using $^{183}$W; all W isotopic ratios have been corrected for these effects (Supplementary Tables S1 and S2; Cook and Schönbächler, 2016). After correction for NFS effects, the $\varepsilon^{182}$W (6/3) values overlap with those obtained with the $^{186}$W/$^{184}$W normalisation scheme (Table 2). Furthermore, all $\varepsilon^{184}$W (6/3) data are within uncertainty of the terrestrial bracketing standard (see Supplementary Material) and indicate that the IAB irons are not characterized by nucleosynthetic variations in W isotopes (e.g., Kruijer et al., 2013a; Kruijer et al., 2017). Thus, no corrections are needed for such effects on $\varepsilon^{182}$W.

## 5. Discussion

### 5.1. Origin of Pt isotope variations

Platinum isotope variations can have two origins. Firstly, variations could reflect the heterogeneous distribution of isotopes formed in specific stellar environments (nucleosynthetic variations) and secondly, isotope ratios can be altered due to exposure to GCR. Nucleosynthetic isotope variations are reported in iron meteorites for elements including Mo (Burkhardt et al., 2011; Dauphas et al., 2002; Worsham et al., 2017), Ru (Chen et al., 2010; Fischer-Gödde et al., 2015) and Pd (Ek et al., 2017; Mayer et al., 2015), but have not previously been recognised for Pt (Hunt et al., 2017b; Kruijer et al., 2014a, 2013a; Peters et al., 2015; Wittig et al., 2013). The low-abundance isotope $^{192}$Pt (0.79%) is the only major $s$-process nuclide of Pt, whereas the more abundant isotopes $^{194}$Pt, $^{195}$Pt, $^{196}$Pt and $^{198}$Pt are all almost uniquely produced by the $r$-process (Bisterzo et al., 2011). Therefore, variable mixing of different nucleosynthetic components should result in $^{192}$Pt variations relative to the other Pt isotopes. Four of the IAB irons (Cranbourne, Livingstone, Magura, Odessa) display well-resolved Pt isotope excesses (Table 2; Fig. 1). However, these $\varepsilon^{192}$Pt excesses correlate with positive shifts in $\varepsilon^{194}$Pt and $\varepsilon^{196}$Pt and therefore the Pt isotope variations do not follow the predicted nucleosynthetic pattern. Additionally, since some samples are identical to the terrestrial standard for all Pt isotope ratios, this provides further strong evidence for the absence of nucleosynthetic variations.

Irradiation of a meteoroid by GCR can readily alter Pt isotope ratios and expected neutron capture effects were modelled by Leya and Masarik (2013). Excesses in $\varepsilon^{196}$Pt are generated by neutron capture on $^{195}$Pt, creating $^{196}$Pt, and are dependent on exposure time, radius, depth within the meteoroid and matrix composition (Leya and Masarik, 2013). Platinum-192 is produced by neutron capture on the relatively abundant $^{191}$Ir (37.3%), followed by $\beta$-decay. Therefore, $\varepsilon^{192}$Pt shifts due to GCR exposure are also highly dependent on the Ir/Pt ratio of the meteorite (Leya and Masarik, 2013). The IAB irons studied here have Ir/Pt ratios of $\sim$0.4, except Livingstone (Ir/Pt $= 0.17$; Table 1; Wasson and Kallemeyn, 2002).

The Pt isotope compositions of the IAB irons lie within uncertainty of the neutron capture model trends for their Ir/Pt ratios (Fig. 1). The Pt isotope deviations of Cranbourne, Livingstone, Magura and Odessa from the terrestrial value fit the model predictions well and can be accounted for solely by GCR exposure, with Odessa being the most strongly exposed of the meteorites studied here. The data for Caddo County, Canyon Diablo and Toluca also fall on the model trend but overlap with the terrestrial standard value, signifying that these samples were only weakly-exposed to GCR. The good fit of our Pt isotope data to the neutron capture model also confirms that nucleosynthetic Pt isotope variations are absent, in agreement with previous studies (Hunt et al., 2017b; Kruijer et al., 2014a, 2013a; Peters et al., 2015; Wittig et al., 2013). This demonstrates that $s$- and $r$-process Pt isotopes were distributed homogeneously in the solar nebula in the region sampled by iron meteorites and the Earth.

### 5.2. Pre-exposure $\varepsilon^{182}$W using combined Pt–W systematics

In an $\varepsilon^{182}$W (6/4) vs. $\varepsilon^{196}$Pt diagram, the data fall on a well-defined linear regression with a slope of $-1.46 \pm 0.34$ (Fig. 2a). The correlation between $\varepsilon^{182}$W (6/4) and $\varepsilon^{192}$Pt is dependent on the Ir/Pt of the samples, and is therefore subject to scatter (Fig. 2b). Normalisation of all samples to Ir/Pt $= 0.4$ ($\varepsilon^{192}$Pt$_{norm}$) reduces this scatter and yields a correlation with a slope of $-0.06 \pm 0.03$. These trends are predicted by the neutron capture models of Leya and Masarik (2013). The neutron capture model trends and the regression through our data mostly overlap within uncertainty (Fig. 2). The model trends, however, feature steeper slopes such that they overestimate the GCR-induced effects on $\varepsilon^{182}$W at more extreme $\varepsilon^{196}$Pt values (Fig. 2a). This was previously observed for IVB irons (Kruijer et al., 2013a; Wittig et al., 2013), and may be due to variations in the compositions of the samples, which slightly alters the neutron energy spectrum and is not accounted for in the modelling (Kruijer et al., 2013a, 2013b). Alternatively, it may reflect the under-production of Pt relative to W in the GCR model (Hunt et al., 2017b). Regardless, the empirically determined slope between $\varepsilon^{182}$W (6/4) and $\varepsilon^{196}$Pt identified here ($-1.46 \pm 0.34$) agrees well with those defined previously ($-1.32 \pm 0.11$, $n = 5$; Kruijer et al., 2014a), and further demonstrates the robustness and usefulness of Pt isotopes as a neutron dose proxy.

Following the approach adopted previously (Kruijer et al., 2014a, 2013a; Wittig et al., 2013), we estimated the $\varepsilon^{182}$W values of our samples before GCR exposure, which in turn can be used to constrain the timing of the last metal–silicate equilibration on the IAB parent body. Using this method, the intercepts of the regressions through the IAB samples where both $\varepsilon^{192}$Pt$_{norm}$ and $\varepsilon^{196}$Pt are 0 yield the $\varepsilon^{182}$W value before GCR exposure. These pre-GCR $\varepsilon^{182}$W (6/4) values are $-2.93 \pm 0.15$ and $-2.89 \pm 0.07$ when correlated against $\varepsilon^{192}$Pt$_{norm}$ and $\varepsilon^{196}$Pt, respectively (Fig. 2). Pre-GCR $\varepsilon^{182}$W (6/3) calculated using $\varepsilon^{192}$Pt$_{norm}$ and $\varepsilon^{196}$Pt yields $-2.93 \pm 0.19$ and $-2.89 \pm 0.15$, respectively. A weighted average of the four pre-GCR $\varepsilon^{182}$W values combines the intercepts defined by the four regressions and yields a single pre-GCR exposure $\varepsilon^{182}$W of $-2.90 \pm 0.06$ for the investigated IAB irons. Based on this value, the timing of metal–silicate differentiation can be calculated relative to CAI formation using the following equation:

$$\Delta t_{CAI} = -\frac{1}{\lambda} \ln\left(\frac{\varepsilon^{182}W_{pre-GCR} - \varepsilon^{182}W_{chondrite}}{\varepsilon^{182}W_{CAI} - \varepsilon^{182}W_{chondrite}}\right) \quad (1)$$

where $\varepsilon^{182}$W$_{pre-GCR}$ is the pre-GCR exposure $\varepsilon^{182}$W calculated for the IAB irons, $\varepsilon^{182}$W$_{chondrite}$ is the composition of carbonaceous chondrites ($-1.9 \pm 0.1$; Kleine et al., 2004), and $\varepsilon^{182}$W$_{CAI}$ is the solar system initial determined from CAIs ($-3.49 \pm 0.07$;

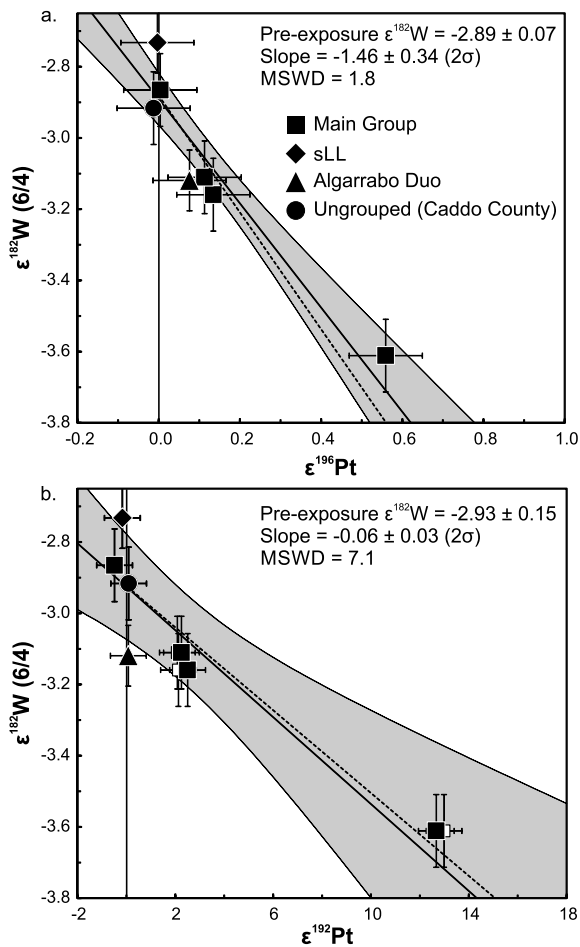

**Fig. 2.** (a) Correlation of $\varepsilon^{182}$W (6/4) with $\varepsilon^{196}$Pt, resulting in a pre-GCR exposure $\varepsilon^{182}$W of $-2.89 \pm 0.07$ and (b) correlation of $\varepsilon^{182}$W (6/4) with $\varepsilon^{192}$Pt, resulting a pre-GCR exposure $\varepsilon^{182}$W of $-2.93 \pm 0.15$ against $\varepsilon^{192}$Pt$_{norm}$. Data in Fig. 2b were normalized to Ir/Pt = 0.4 (closed symbols). Where normalisation resulted in a difference to $\varepsilon^{192}$Pt, the original data are shown as open symbols. Regression lines and error envelopes were calculated using ISOPLOT (Ludwig, 2003). Also displayed are GCR model calculations from Leya and Masarik (2013) for an exposure time of 1000 Ma, Re/W = 0.25, Os/W = 3.125 and Ir/Pt = 0.4 (dashed line). The GCR model lies within error of the regression lines, however, Fig. 2a shows that the model tends towards more negative $\varepsilon^{182}$W for higher $\varepsilon^{196}$Pt. Uncertainties on $\varepsilon^{i}$Pt are based on the North Chile in-house standard (2 S.D.); uncertainties on $\varepsilon^{i}$W are based on the W bracketing standard (NIST SRM 3163) or the long-term reproducibility defined by repeated measurements of SRM 129c (Cook and Schönbächler, 2016; see Table 2 for further details), whichever is larger.

Burkhardt and Schönbächler, 2015; Kruijer et al., 2014b). This equation yields a model age for Hf–W fractionation, assuming a previous uniform reservoir with a chondritic Hf/W ratio. To facilitate comparison and consistency with previous studies (i.e., Kruijer et al., 2014a, Schulz et al., 2012, 2009; Worsham et al., 2017), the uncertainties on the CAI and CI values are not taken into account, both for our new data and for the data from previous studies recalculated relative to the latest value for CAI. Using our new weighted average ($\varepsilon^{182}$W = $-2.90 \pm 0.06$), equation (1) yields a time of metal–silicate separation in the IAB parent body of $6.0 \pm 0.8$ Ma after CAI formation. Alternatively, our best single regression ($-2.89 \pm 0.07$ for $\varepsilon^{182}$W (6/4) vs. $\varepsilon^{196}$Pt) defines a nearly-identical metal–silicate separation event at $6.1 \pm 0.9$ Ma after CAI formation.

A single metal–silicate separation age for all IABs investigated here may be surprising for the following reason. Of the seven samples analyzed, only four (Canyon Diablo, Cranbourne, Magura and Odessa) belong to the IAB main group. Toluca belongs to the sLL sub-group, Livingstone (TN) is part of the Algarrabo duo, and Caddo County is ungrouped. These different groups are suggested to have formed separately through independent impacts into the parent body, forming distinct melt pools (Choi et al., 1995; Wasson and Kallemeyn, 2002; Worsham et al., 2016, 2017). Therefore, these melt pools may have different metal–silicate separation ages, depending on the timing of the melt-generating impact. Samples from all four groups, however, fall on the same regression for $\varepsilon^{182}$W vs. $\varepsilon^{19i}$Pt. A regression through the MG-IAB irons alone yields a pre-exposure $\varepsilon^{182}$W of $-2.95 \pm 0.04$, which is identical within uncertainty to the pre-exposure $\varepsilon^{182}$W value obtained from the regression of all seven IAB samples ($\varepsilon^{182}$W = $-2.90 \pm 0.06$). Moreover, Caddo County, Toluca and Livingstone are not resolvable from the members of the main group IAB irons (Fig. 2). This further corroborates that the MG irons and sub-groups investigated here experienced metal–silicate differentiation contemporaneously. However, there is some scatter within our dataset. A study with higher precision would be needed to clarify whether this is solely due to analytical scatter or represents temporally distinct melting events on the IAB parent body, particularly between the MG irons and sLL sub-group.

Nucleosynthetic Mo isotope variations indicate that the IAB sub-groups may originate from numerous parent bodies (Worsham et al., 2017). Nucleosynthetic variations have been described between some iron meteorite groups for $\varepsilon^{183}$W (e.g., Burkhardt et al., 2012; Kruijer et al., 2013a, 2017) and this may allow multiple parent bodies for the IAB complex to be distinguished. However, no W nucleosynthetic variations are present between the samples measured here (Table 1), consistent with a single parent body origin. Furthermore, a lack of nucleosynthetic variations in Pd isotopes between all groups studied here (Ek et al., 2016), and a homogeneous Mo isotope composition between the MG and sLL sub-group (Worsham et al., 2017) permit derivation from a single parent body.

### 5.3. Comparison to previous studies

The age of $6.0 \pm 0.8$ Ma after CAI formation is more precise, but in good agreement with several other studies (e.g., Markowski et al., 2006; Qin et al., 2008; Schulz et al., 2009). Moreover, a metal–silicate separation age of $6.0 \pm 0.8$ Ma after CAI agrees well with a postulated silicate melting event recorded in the winonaites at $6.6 \pm 2.9$ Ma (Schulz et al., 2010; recalculated relative to the latest CAI value, Burkhardt and Schönbächler, 2015; Kruijer et al., 2014b), further corroborating the close relationship between IABs and winonaites that is suggested by geochemical and textural similarities, and Mo and O isotopes (e.g., Benedix et al., 2000; Clayton and Mayeda, 1996; Worsham et al., 2017).

Our new age also agrees with the value of $6.9 \pm 0.4$ Ma (recalculated relative to the newest CAI estimate; Burkhardt and Schönbächler, 2015; Kruijer et al., 2014b) determined by Schulz et al. (2012) for the majority of IAB irons studied therein (Fig. 3). However, Schulz et al. (2012) do not obtain a single pre-exposure age for the IAB samples also studied here (e.g., Caddo County). Those authors used a correlation of $\varepsilon^{182}$W with CRE ages (determined using cosmogenic noble gases) defined by the majority of their samples to obtain a pre-exposure $\varepsilon^{182}$W. This approach works to a first order, but our data demonstrate that longer CRE ages do not necessarily lead to a linear increase in $\varepsilon^{196}$Pt (Fig. 4a), and by extension $\varepsilon^{182}$W. This is because cosmogenic noble gases are generally produced by high energy reactions at shallow depths in a meteoroid, whereas W and Pt isotopes undergo lower energy (thermal–epithermal) neutron capture reactions at greater depths (e.g., Kruijer et al., 2013a; Leya and Masarik, 2013; Wittig et al., 2013). Furthermore, the (epi-)thermal neutron fluence also depends on the pre-atmospheric radius and depth of the sample within the meteoroid (Fig. 4b), and this effect is not accounted for by CRE ages. Therefore, CRE ages are suitable to identify unex-posed samples, but do not provide a means of accurate correction of the GCR effects on $\varepsilon^{182}$W.

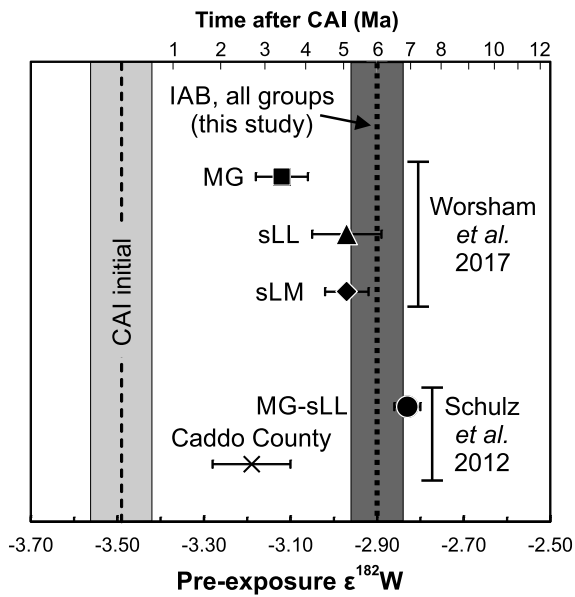

**Fig. 3.** Overview of pre-exposure $\varepsilon^{182}$W for the IAB irons from this study (dark-grey band), compared to data from Schulz et al. (2012) and Worsham et al. (2017), relative to the CAI value (light-grey band; Kruijer et al., 2014b). Our study demonstrates that the wide range of scatter in $\varepsilon^{182}$W for the IABs is partly related to neutron capture and not distinct timings for individual metal–silicate separation events on the parent body.

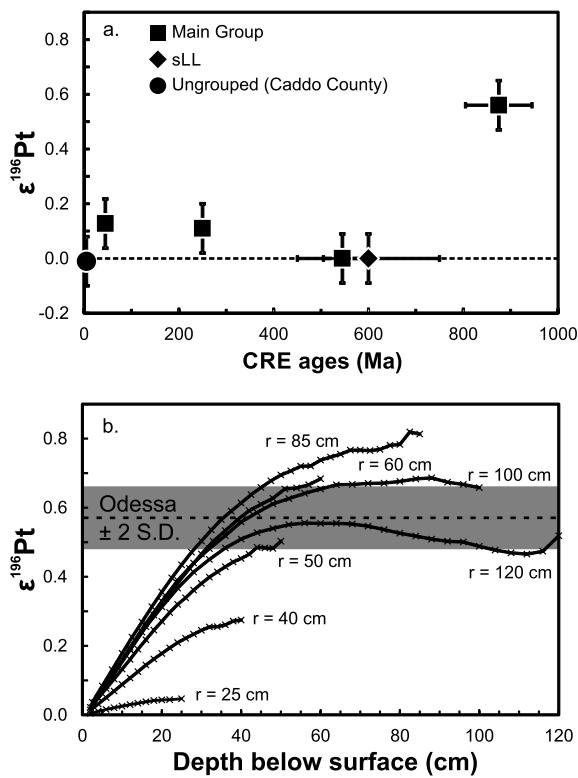

**Fig. 4.** (a) Plot of $\varepsilon^{196}$Pt vs. cosmic ray exposure age. This graph shows no linear increase of $\varepsilon^{196}$Pt with CRE age because $\varepsilon^{196}$Pt is strongly dependent on the depth of the sample below the surface. CRE ages are therefore not suitable to correct for neutron capture effects in iron meteorites. See Table 1 for sources of CRE ages. (b) Expected GCR-induced shift in $\varepsilon^{196}$Pt for meteoroids with pre-atmospheric radii between 25 and 120 cm and an exposure time of 875 Ma (Leya and Masarik, 2013). The offset in $\varepsilon^{196}$Pt is strongly dependent on both the pre-atmospheric radius and the depth of the sample below the surface. Dashed black line and grey shaded area show $\varepsilon^{196}$Pt for Odessa (exposure time = 875 ± 70 Ma; Voshage and Feldmann, 1979).

An age of 6.0 ± 0.8 Ma is also consistent with that determined for the sLL and sLM sub-groups (5.0 ± 1.0 and 5.1 ± 0.6 Ma, respectively; Fig. 3) by Worsham et al. (2017). The segregation of the sLM sub-group at this time adds further evidence for a body-wide event. However, Worsham et al. (2017) determined an earlier age of metal–silicate segregation for the MG (3.4 ± 0.7 Ma). Combining our dataset with the MG of Worsham et al. (2017) yields a pre-exposure $\varepsilon^{182}$W of −3.01 ± 0.08, which is similar to the value deduced from our data only and corresponds to metal–silicate segregation at 4.6 ± 0.9 Myr after CAI. For the further discussion we will use the age derived from our data only because the discrepancy between the two datasets may be partly a result of the method of GCR correction employed by Worsham et al. (2017). For 3 out of the 4 MG samples analysed in that study, the neutron-dose proxy (Os isotopes) and W isotopes were not determined on the same sample aliquot, but on separate digestions of the samples. Models of GCR effects in iron meteorites indicate that moving 2–3 cm deeper into an iron meteorite results in a $\varepsilon^{182}$W (6/4) shift which is double that modelled for $\varepsilon^{189}$Os (Leya and Masarik, 2013). This is supported by correlated Os and W isotope ratios for the IID iron meteorite Carbo and the IVB irons, where series of variably exposed samples display offsets induced by GCR that are greater for $\varepsilon^{182}$W than for $\varepsilon^{189}$Os (Qin et al., 2015; Wittig et al., 2013). An inaccurate GCR correction may therefore induce errors of ∼2–4 ppm to $\varepsilon^{182}$W, assuming aliquots for W and Os analyses are sampled within 3 cm of each other. However, this effect is very unlikely to fully account for the discrepancy between our study and the MG data of Worsham et al. (2017). Nevertheless, it demonstrates the importance of determining W isotopes and a neutron-dose proxy such as Pt or Os isotopes on the same sample aliquot, particularly when high-precision data are required.

Using our approach, all IAB samples define a single value for pre-GCR exposure $\varepsilon^{182}$W. This demonstrates that, within the resolution of this study, the analysed IAB samples experienced metal–silicate separation contemporaneously and several distinct melting events, as previously suggested, cannot be resolved (Schulz et al., 2012; Worsham et al., 2017).

### 5.4. Thermal evolution of the IAB parent asteroid

#### 5.4.1. Hypotheses for IAB parent asteroid evolution

Two hypotheses have been proposed to explain the evolution of the IAB parent body. Local impact-generated melt pools were proposed based on element concentration data of IAB metals, which display only limited fractional crystallisation trends caused by modest separation of solid metal from the remaining metal melt. At the same time, the IAB concentration data show significant scatter and this was accommodated by invoking various impact-related melt pools (Choi et al., 1995; Wasson and Kallemeyn, 2002; Worsham et al., 2016). However, our new data show that the main group IAB irons, multiple sub-groups (sLL and sLM), plus one duo and an ungrouped sample, experienced metal–silicate differentiation contemporaneously. This uniform age argues for a single global-scale metal–silicate separation event. Our data cannot exclude the scenario in which the five groups formed in separate melt pools induced by impacts. Nonetheless, in order to reconcile the coeval metal–silicate separation time of the MG-IAB irons and sub-groups with impact-related heat sources, multiple impacts large enough to induce significant melting would have to occur almost simultaneously, or a single impact generated several different large melt pools. Moreover, it has been argued that impacts alone cannot provide enough heat to melt significant fractions of a parent body, thereby excluding them as a global heat source (Ciesla et al., 2013; Davison et al., 2012).

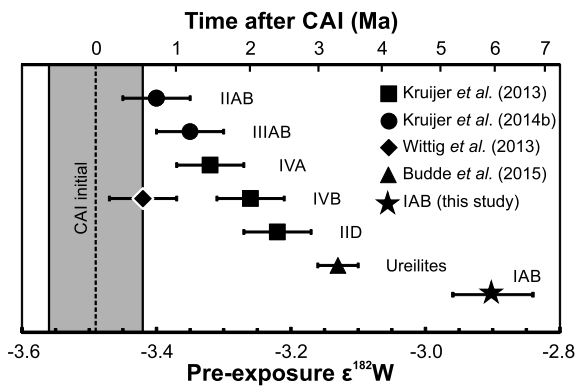

**Fig. 5.** Overview of pre-GCR exposure $\varepsilon^{182}$W and corresponding metal–silicate separation ages, determined by correlation with Pt isotopes for magmatic iron meteorite groups (Kruijer et al., 2014a, 2013a; Wittig et al., 2013) and IAB irons (this study), plus $\varepsilon^{182}$W for weakly-exposed ureilites (Budde et al., 2015). This illustrates the earlier metal–silicate separation age of the magmatic irons compared to primitive achondrites and related parent bodies (ureilites, IAB irons). Grey shaded area denotes CAI and associated uncertainty (Kruijer et al., 2014b).

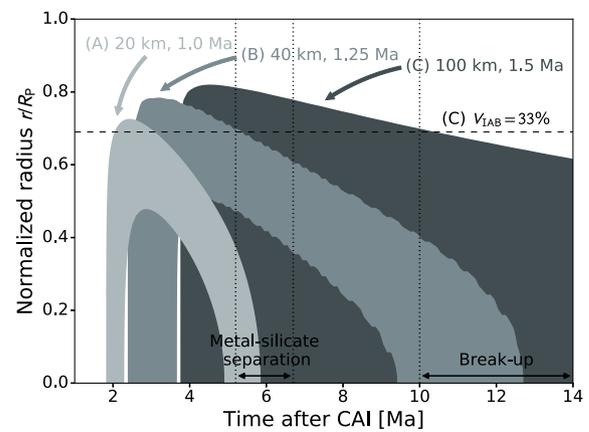

**Fig. 6.** Time evolution of the interior region experiencing temperatures in agreement with temperature constraints (1240 K – upper bound of each band, and 1470 K – lower bound of each band) until 14 Ma after CAI. Any model eligible as the IAB parent asteroid configuration must exhibit the above temperatures from the moment of metal–silicate separation (6.0 ± 0.8 Ma) until the time of catastrophic break-up (10–14 Ma after CAI). The radii and formation times for models (A), (B) and (C) are given in the figure. Model (A) cooled too fast to sustain sufficiently high temperatures until break-up, while model (B) reached temperatures above 1470 K before metal–silicate separation and was thus discarded. Model (C) and all 2D models formed at 1.5 Ma with radii larger than 60 km experienced internal temperatures within the correct temperature bracket between the estimated times for metal–silicate separation and catastrophic break-up, and featured up to 60% as the eligible volume. The dashed line, (C) $V_{IAB}$, illustrates this volume for model (C).

Benedix et al. (2000) proposed an alternative model, which argues that the decay of short-lived radionuclides produced enough heat for incipient partial melting. This was followed by a catastrophic impact causing break-up and reassembly of the parent body, which led to the juxtaposition of lithologies such as unmelted, chondritic silicates mixed with molten metal. In the context of their model, they put forward that metamorphism, partial melting and incomplete separation of metallic melt from the residue occurred to explain the element concentrations of the metals and silicate inclusions. At the same time, it is likely that some of the metal, in particular incompletely extracted metal closer to the surface, had already solidified by the time of impact and reassembly. These metal pools could have been locally remelted by the disrupting impact, adding to the scatter in the element concentration data of the IABs. Theis et al. (2013) proposed a time interval for the catastrophic break-up and reassembly of the parent body based on the $^{107}$Pd–$^{107}$Ag system and other chronometers that all record a heating event around 10 to 14 Ma after CAI. A model of internal heating plus later break-up and reassembly may provide an explanation for the differentiation event recorded by the IAB groups studied here. However, the age of the IABs is significantly younger than those of the magmatic iron meteorites, which likely differentiated within the first 1–4 Ma after CAI formation as a result of internal heating (Fig. 5; Kruijer et al., 2014a, 2013a; Wittig et al., 2013). These different timescales merit an explanation.

#### 5.4.2. Numerical models of IAB asteroid thermal evolution

The new and precise metal–silicate separation age (6.0 ± 0.8 Ma after CAI) provides constraints on the accretion time and subsequent thermal evolution of the IAB parent body. Thermal models incorporating internal heating by radioactive nuclides ($^{26}$Al and $^{60}$Fe) can help to constrain whether it is possible to model the features described for the IAB parent body via internal heating, and if so, to constrain the relationship between the timing of accretion and metal–silicate separation as well as parent body size.

Brecciated silicate clasts are present in many IAB samples (Table 1; e.g., Benedix et al., 2000, 2005; Buchwald, 1975; Ruzicka and Hutson, 2010) and these textures, combined with peak temperatures obtained from both silicate clasts and the related winonaite meteorites set upper and lower bound temperatures for the timing of metal–silicate separation and break-up. Textures in both the IAB irons and winonaites indicate the presence of molten metal, which implies that large portions of the parent body reached at least the cotectic melting point for the Fe, Ni–FeS system (∼1220–1261 K; Kubaschewski, 1982; Kullerud, 1963) and sets a minimum temperature boundary. Additionally, the majority of the body reached a maximum temperature of ∼1470 K, based on peak temperatures obtained from both silicate clasts and winonaite meteorites (Benedix et al., 1998, 2000) and winonaite trace-element geochemistry (Hunt et al., 2017a). A body in this temperature range cannot fully differentiate, thus allowing for the preservation of the range of compositions observed in IAB irons and the winonaites. This includes both the residues of silicate melting and the melts themselves, plus abundant chondritic lithologies (e.g., Benedix et al., 2000; Hunt et al., 2017a; Ruzicka and Hutson, 2010). However, partial core formation could set in via incomplete percolative flow through the silicate matrix, which would leave portions of the metal in the mantle (Bagdassarov et al., 2009; Cerantola et al., 2015). An additional constraint is provided by the mixing of molten metal with cool silicates during the parent body break-up and reassembly at 10 to 14 Ma after CAI (Benedix et al., 2000; Theis et al., 2013). This implies that molten metal is still present at this time, at some depth in the pre-impact IAB parent body and requires a portion of the body to remain above ∼1240 K. Our new study dates metal segregation to 6.0 ± 0.8 Ma, and hence, metal must stay molten for at least another 4 Ma after metal–silicate separation. We used these constraints in our thermal models to identify planetesimals that satisfy these temperature criteria at the time of metal–silicate separation and break-up (Fig. 6).

Depending on initial radius and formation time, and adopting the assumption of an initial homogeneous distribution of $^{26}$Al in the protoplanetary disk, the numerical models display different evolutions. Larger radii and earlier formation times lead to higher internal temperatures. The radius primarily controls the cooling timescale, such that small bodies cool quicker than larger ones (Fig. 6). Therefore, larger bodies sustained specific temperature regimes at a specific radius for longer than smaller bodies. Deviations between 2D and 3D models resulted from changed surface-to-volume ratios in the 3D runs, which enabled these 3D models to cool more efficiently. Modelling a large number of bodies with varying radii and formation times, combined with the constraints from metal–silicate separation age and break-up time (see above), limits the possible formation time and radius of the IAB asteroid parent body (Fig. 7). Many modelled parent bodies show only very narrow regions (eligible volume, $V_{IAB}$) within the correct temperature regime or do not sustain temperatures above 1240 K until 10 Ma after CAI or later (Figs. 6 and 7); these bodies were discarded as potential IAB parent bodies.

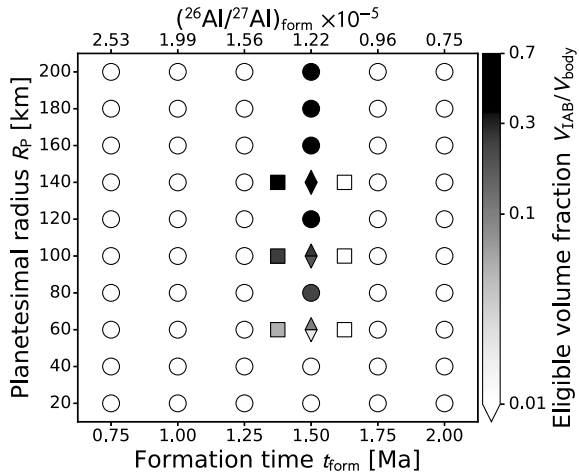

**Fig. 7.** Evaluated parameter space showing 2D (circles) and 3D (squares) numerical simulations of the interior thermal evolution of potential IAB parent asteroid bodies. Models marked with a diamond were run using both geometries, with the upper part of the diamond indicating the obtained value for the 2D runs and the lower part representing the 3D results. Gray intensity scales display the fraction of eligible volume at 10 Ma for each model (which for 2D models was interpolated to a 3D volume) satisfying the temperature constraints at the estimated times for metal–silicate separation and catastrophic break-up (cf., Fig. 6 for three showcase examples). Only bodies formed at $1.4 \pm 0.1$ Ma (or equivalently, $^{26}$Al/$^{27}$Al $\sim(1.35 \pm 0.13) \times 10^{-5}$) robustly satisfy the temperature criteria derived for the IAB parent asteroid body.

Model planetesimals with radii larger than 60 km and formation times $t_{form} = 1.4 \pm 0.1$ Ma after CAI satisfy the temperature constraints for a significant portion of their interior, which qualifies them as eligible parent asteroids for the IAB meteorites. Internal heating is therefore a viable mechanism for the formation of the IAB irons. These models reached the appropriate temperature regime throughout major parts of their interiors for more than 10 Ma after their formation. Additionally, some 2D models with an earlier accretion time ($t_{form} = 1.25$ Ma) and radius $60 \text{ km} \leq R_P \leq 120$ km showed minor eligible volume fractions which could be the source of IAB meteorites. However, these experienced temperatures above 1470 K before the recorded time for metal–silicate separation, and thus should have undergone complete metal–silicate separation accompanied by closure of the $^{182}$Hf–$^{182}$W system. These models were therefore discarded as viable solutions.

For all models, the time interval between Fe, Ni–FeS melting within the body and the closure of the $^{182}$Hf–$^{182}$W system was approximately $\Delta\tau_{seg} = 2$–3 Ma (Fig. 6). Liquid metal can percolate approximately $\Delta x = \Delta\tau_{seg} k \Delta\rho g/\mu \approx 500$–1000 m during this time interval assuming the following: perfect Darcy flow with constant melt fraction, no compaction (for parameter references see Bagdassarov et al., 2009), a permeability of $k = 10^{-15}$ m$^2$ at 10% melt fraction, Fe–S shear viscosity $\mu$ of $10^{-2}$ Pa s, and FeS melt to solid peridotite density contrast of $\Delta\rho = 1900$ kg m$^{-3}$ at 50 km distance from the center of the planetesimal. This model and the resulting liquid metal percolation speed allows metal–silicate segregation via interconnected percolation networks in the appropriate time frame of 2–3 Ma until closure of the $^{182}$Hf–$^{182}$W chronometer. Furthermore, $^{182}$W can diffuse from silicates to metal during cooling with diffusion length scales per time unit depending on temperature (Quitté et al., 2005). Therefore, if metal existed as fine veins and small pools in the mantle in the initial stage of percolation, the $^{182}$Hf–$^{182}$W system may not have fully closed to re-equilibration, thereby explaining the time gap between the models (Fig. 6) and the recorded metal–silicate separation age. Variable closure times for the metal pools may also explain the scatter between the MG and sLL noted in our dataset (section 5.2).

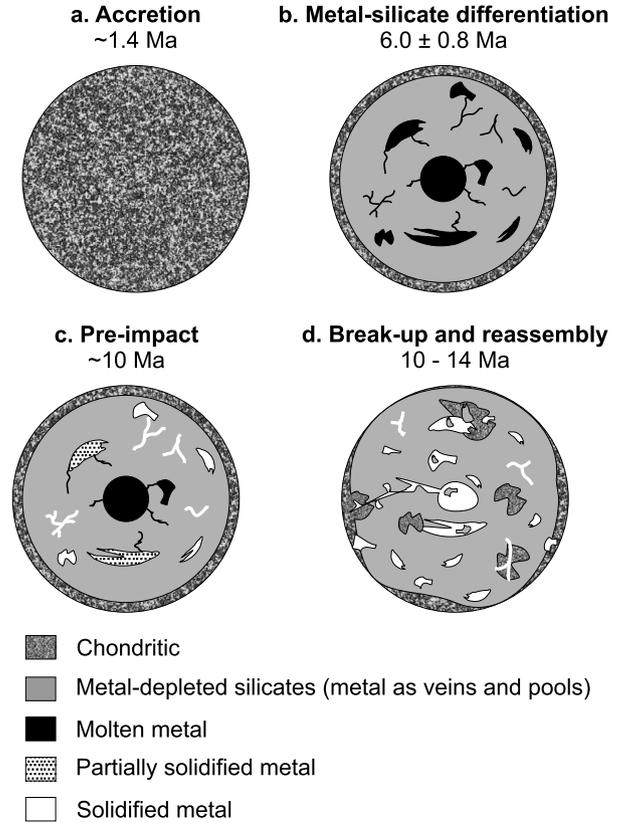

**Fig. 8.** Overview of the temporal evolution of the IAB parent body. (a) Accretion of chondritic material at $\sim$1.4 Ma after CAI. Parent body has a radius of 60 km or greater. (b) Metal–silicate differentiation at $6.0 \pm 0.8$ Ma. This is a result of internal heating by short-lived radionuclides. Molten metal was not efficiently extracted into a core, and also existed as pools and veins. (c) Prior to the catastrophic impact, the body began to cool and metal pools and veins in the outer parts of the body may have begun to solidify. However, liquid metal was still present. (d) Impact-related break-up and reassembly of the parent body. This event occurred between 10 and 14 Ma after CAI (Theis et al., 2013) and led to extensive mixing of lithologies in the parent body.

However, the $^{182}$Hf–$^{182}$W system must essentially be closed by $\sim$6 Ma in order to preserve the measured W isotope ratios reported here. This may happen as the liquid metal coalesces into larger ponds and diffusion from silicates becomes negligible. Additional diffusion of W may have occurred during the break-up and reassembly stage when the silicate inclusions were mixed into the metal. However, the initial cooling to below the Fe, Ni–FeS solidus ($\sim$1220 K) must have been very fast to keep the inclusions in suspension in the liquid metal (e.g., Wasson and Kallemeyn, 2002). Additionally, W diffusion in silicates is very slow ($10^{-17}$ cm$^2$ s$^{-1}$ at $\sim$1423 K; Quitté et al., 2005). Hence, diffusion may have led to some small-scale metamorphic resetting of $^{182}$Hf–$^{182}$W ages in silicate inclusions (Schulz et al., 2009), but was not sufficient to reset the bulk metal ages. Furthermore, Schulz et al. (2012) noted that both bulk metal and metal entrained in silicate inclusions in the Landes meteorite (IAB-MG) showed the same W isotope signatures.

To summarize, in an environment with homogeneous, canonical $^{26}$Al/$^{27}$Al, the IAB parent body accreted around $\sim$1.4 Ma and likely had a radius $\geq$60 km (Fig. 8a), as constrained by our numerical thermal models. Members of the IAB main group and the sub-groups investigated here underwent metal–silicate separation at 6.0 ± 0.8 Ma (Fig. 8b). This metal–silicate separation event is late in comparison to the magmatic iron meteorite parent bodies but still early enough to be the result of internal heating by means of the decay of short-lived radionuclides. As shown by previous modelling, a primordial chondritic shell can survive in the upper layers of a planetesimal (Lichtenberg et al., 2016). Therefore a body consisting of metamorphosed but broadly chondritic (winonaite-like) material, with pools and veins of molten metal inefficiently extracted into a core, is feasible. Lithologies present in the IAB irons show unambiguous evidence of mixing by impact, which can be explained by a later catastrophic break-up and reassembly of the IAB parent body (Benedix et al., 2000). Thermal models indicate that certain parts of the parent body sustain temperatures high enough to keep liquid metal until the time it is mixed with solid silicates during the reassembly event (Fig. 8c), although metal pools closer to the surface may have been crystallised or partially crystallised. This catastrophic break-up event is dated to 10–14 Ma after CAI formation (Theis et al., 2013). Once reassembled, the interior portions of the body were scrambled, mixing cooler silicates into molten metal (Fig. 8d).

*5.4.3. Implications for the timing of core formation in planetesimals*

There are only a few other parent bodies known to have experienced delayed metal–silicate differentiation. Among these are ureilites, with a postulated accretion at ∼1.6 Ma and a metal–silicate separation age of 3.3 ± 0.7 Ma after CAI (Budde et al., 2015), which is slightly older than that calculated here for the IAB irons (Fig. 5). Additionally, the parent body for the acapulcoite–lodranites underwent metal–silicate separation at 4.7 ± 1.1 Ma after CAI formation (Schulz et al., 2010; recalculated to the latest CAI estimate, Burkhardt and Schönbächler, 2015; Kruijer et al., 2014b), thus in the same time period as the IAB parent body. Both the ureilite and acapulcoite–lodranite parent bodies are also interpreted to have undergone incomplete partial melting and core-mantle differentiation as a result of subdued internal heating by radioactive nuclides (Budde et al., 2015; Golabek et al., 2014).

The thermal models used to predict the timing of accretion assume a homogeneous, canonical $^{26}$Al/$^{27}$Al (∼5.25 × 10$^{-5}$; Kita et al., 2013) at the time of planetesimal formation. Some evidence suggests that this assumption may be incorrect and that $^{26}$Al was heterogeneously distributed in the early solar system (Larsen et al., 2011; Schiller et al., 2015). The IAB parent asteroid may still form and partially differentiate via internal heating in a low $^{26}$Al environment, but the timing of accretion cannot be as accurately constrained. Under such conditions, the parent bodies displayed in Fig. 6 (which are associated with specific initial $^{26}$Al/$^{27}$Al that determines the temperature history of the body; see section 3.2) can no longer be linked to specific formation times. Generally, the bodies are shifted to earlier formation times, and the IAB parent body had to accrete before ∼1.4 Ma (as in previous models, see section 5.4.2) otherwise formation in an environment with supra-canonical $^{26}$Al is implied. One consequence of earlier formation is the lower eligible volume fraction ($V_{IAB}$) that remains in the correct temperature range at ∼10 Ma after CAI for each planetesimal model, because of the prolonged cooling experienced by the bodies. However, this does not qualitatively change our conclusions. For the case of heterogeneous $^{26}$Al distribution, comparisons with other early-formed bodies are not straightforward; however, one implication would be that if $^{26}$Al was heterogeneously distributed in the early solar system, then the primitive achondrites could form at the same time as the magmatic iron parent bodies, but in a lower $^{26}$Al environment. In this scenario, the lower radiogenic heat budget available implies that the IAB-winonaite, acapulcoite–lodranite, and ureilite parent bodies were never able to fully differentiate, despite their postulated earlier accretion times.

## 6. Conclusions

Tungsten isotopes are affected by neutron capture when exposed to GCR, altering their isotopic ratios. This hampers the use of the $^{182}$Hf–$^{182}$W chronometer and can lead to erroneous ages for metal–silicate separation. Here, we use Pt isotopes as a neutron-dose proxy to correct for W isotope shifts in IAB iron meteorites. Platinum and W isotope data for the IAB irons fall along the trends of model calculations for expected neutron capture effects, and Pt isotope variations are wholly due to exposure to GCR. By correlating Pt and W isotope ratios, a single pre-exposure $\varepsilon^{182}$W value of −2.90 ± 0.06 can be derived for the IAB irons. This corresponds to a metal–silicate separation age of 6.0 ± 0.8 Ma after CAI for the IAB parent body.

We propose that this event is still early enough such that short-lived radionuclides caused metal–silicate separation. Numerical modelling indicates that this likely occurred in a parent body with a radius of 60 km or greater, provided that accretion occurred around 1.4 ± 0.1 Ma after CAI formation. This resulted in a partially differentiated body where metal was not fully segregated into the core, but formed metal veins and pools. A later impact caused a major disruption on the parent body between 10 and 14 Ma after CAI formation. This event fragmented the IAB parent body, which was subsequently reassembled, leading to the mixing of different lithologies. Our results confirm that the parent bodies of non-magmatic iron meteorites and primitive achondrites accreted and differentiated later than those of magmatic iron meteorites. Alternatively, in the scenario where $^{26}$Al was distributed heterogeneously in our solar system, our models indicate that primitive achondrite and iron meteorite parent bodies may have accreted at the same time. However, the primitive achondrite parent bodies were not able to fully differentiate due to a lower initial $^{26}$Al/$^{27}$Al ratio and therefore a reduced radiogenic heat budget.

## Acknowledgements

This work was supported by the European Research Council under the European Union's Seventh Framework Programme (FP7/2007–2013)/ERC Grant agreement No. [279779]. T.L. was supported by ETH Research Grant ETH-17 13-1. The numerical simulations in this work were performed on the EULER high performance computing cluster of ETH Zürich. T.L. and G.J.G. thank Taras V. Gerya for providing the I2/I3MART code family. Parts of this work have been carried out within the framework of the National Center for Competence in Research PlanetS supported by the Swiss National Science Foundation. We wish to thank Frederic Moynier for editorial handling of this manuscript, and Peter Sprung and a second reviewer for their helpful insights. A.C.H. additionally thanks Gretchen Benedix for many discussions on the IABs and winonaites. We would like to thank the Smithsonian Institution (Julie Hoskin) for the loan of Livingstone (TN).

## Appendix A. Supplementary material

Supplementary material related to this article can be found online at https://doi.org/10.1016/j.epsl.2017.11.034.